\documentclass[aps,prl,showpacs,floatfix,twocolumn,amsmath,amssymb,preprintnumbers]{revtex4-1}
\usepackage{mathrsfs}
\usepackage[figuresright]{rotating}
\usepackage{amsmath}
\usepackage{amssymb}
\usepackage{graphicx}
\usepackage{color}
\usepackage{dcolumn}
\usepackage{bm}
\usepackage[breaklinks=true,colorlinks=true,linkcolor=blue,urlcolor=blue,citecolor=blue]{hyperref}

\usepackage{soul}

\makeatletter
\newcommand{\rmnum}[1]{\romannumeral #1}
\newcommand{\Rmnum}[1]{\expandafter\@slowromancap\romannumeral #1@}
\makeatother

\begin{document}

\title{Anisotropic magnetocrystalline coupling of the skyrmion lattice in MnSi}

\author{Yongkang Luo$^{1}$}
\email[]{mpzslyk@gmail.com}
\author{Shi-Zeng Lin$^{1}$}
\author{D. M. Fobes$^{1}$}
\author{Zhiqi Liu$^{1}$}
\author{E. D. Bauer$^{1}$}
\author{J. B. Betts$^{1}$}
\author{A. Migliori$^{1}$}
\author{J. D. Thompson$^{1}$}
\author{M. Janoschek$^{1}$}
\author{B. Maiorov$^{1}$}
\email[]{maiorov@lanl.gov}
\address{$^1$Los Alamos National Laboratory, Los Alamos, New Mexico 87545, USA.}

\date{\today}

\begin{abstract}

We investigate the anisotropic nature of magneto-crystalline coupling between the crystallographic and skyrmion crystal (SKX) lattices in the chiral magnet MnSi by magnetic field-angle resolved resonant ultrasound spectroscopy. Abrupt changes are observed in the elastic moduli and attenuation when the magnetic field is parallel to the [011] crystallographic direction. These observations are interpreted in a phenomenological
Ginzburg-Landau theory that identifies switching of the SKX orientation to be the result of an anisotropic magneto-crystalline coupling potential.
Our work sheds new light on the nature of magneto-crystalline coupling potential relevant to future spintronic applications.

\end{abstract}


\maketitle


Future computation requires not only high-speed information transport but also low-energy, ultra-stable and high-density data storage. Magnetic skyrmions, topologically protected swirling spin textures, have recently been established as a platform for easy spin manipulation and, in turn, are highly promising for these next-generation applications. Skyrmions carry a topological charge $N$=$\frac{1}{4\pi}$$\int$$\mathbf{n}$$\cdot(\partial_x\mathbf{n}$$\times$$\partial_y\mathbf{n})d^2\mathbf{r}$=$\pm1$ \cite{Nagaosa-SKTopog}, that remains unchanged under continuous transformation of their magnetic configuration, and thus provides the basis for ultra-stable memories. Here $\mathbf{n}$ is a unit vector that denotes the direction of the magnetic moments. Skyrmions form a periodic spin crystal by packing into a hexagonal lattice called a skyrmion crystal (SKX). The SKX only pins weakly to defects\cite{Nii-MnSiElastic} and, in turn, can be driven by electric current density as low as 10$^6$ A/m$^2$, five orders of magnitude smaller than in conventional spintronic materials based on domain-wall motion\cite{Jonietz-MnSiCurrent,Neubauer-MnSiTHE,Schulz-MnSiTHE}.

The continuously growing list of non-centrosymmetric cubic materials\cite{Muhlbauer-MnSiSKX2009,Munzer-FeSi_CoSKX,Pfleiderer-B20SKX,Seki-Cu2OSeO3SKX,Tokunaga-CoMnZnSKX,Kezsmarki-GaV4S8SKX} as well as thin films\cite{Yu-FeSi_CoFilmSKX,Yu-FeGeFilmSKX,Heinze-FeIrSKX,JiangWJ-Science2015} that exhibit skyrmions of varying sizes (0.5-200 nm) further suggests that SKX may be tailored for specific memory applications. For example, larger skyrmions may couple less to the underlying crystallographic lattice and are candidates for fast memories, whereas sub-nm skyrmions may offer unprecedented memory densities. Thus, the exact nature of the magneto-crystalline coupling between the SKX lattice and the underlying crystallographic lattice is a critical issue. Here we employ systematic field-angle resolved resonant ultrasound spectroscopy (RUS) measurements on the prototypical skyrmion compound MnSi, complemented with theoretical calculations, to provide new insights into the anisotropic nature of the magneto-crystalline coupling potential in cubic skyrmion materials.

\begin{figure}[t]
\vspace*{-20pt}
\hspace*{-10pt}
\includegraphics[width=9.5cm]{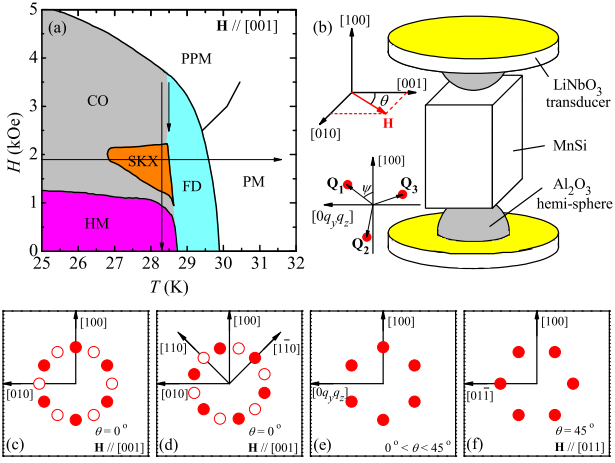}
\vspace*{-18pt}
\caption{\label{Fig1} (a) Phase diagram of MnSi for $\mathbf{H}$$\parallel$[001]. The arrows manifest the trajectories of field- and temperature sweeps with field-rotation. The abbreviations are: SKX=Skyrmion crystal, CO=Conical, HM=Helimagnetic, (P)PM=(Polarized) Paramagnetic, and FD=fluctuations-disordered. (b) Schematic diagram of RUS experimental setup for field rotation. (c-f) Schematic representation of possible SKX orientations in reciprocal space for different field directions. Solid and open circles denote two degenerate domains, respectively.}
\end{figure}

The magnetic phase diagram of MnSi is representative of the class of cubic B20 materials that include metallic, semiconducting and insulating compounds, all of which show SKX phases\cite{Muhlbauer-MnSiSKX2009,Munzer-FeSi_CoSKX,Pfleiderer-B20SKX}. We selected MnSi, for which the first SKX was reported\cite{Muhlbauer-MnSiSKX2009}, for our study because of the extensive knowledge available about its magnetic structure and interactions. The ground state of MnSi is a long-pitch helimagnetic (HM) order arising as a consequence of competing ferromagnetic exchange and Dzyaloshinsky-Moriya interactions (DMI)\cite{Landau-CTP}. Here the helix propagates parallel to the [111] crystallographic axis. Application of an external magnetic field ($\mathbf{H}$) initially rotates the helix towards the field axis and then polarizes the spins, causing sequential HM-Conical (CO)-Polarized paramagnetic (PPM) phase transitions [see Fig.~\ref{Fig1}(a)]. The SKX phase emerges for moderate $H$ just below the HM transition temperature $T_c$.

The hexagonal SKX can be regarded as the superposition of three magnetic spirals with propagation vectors $\mathbf{Q}$ rotated by $120^\circ$ in the same plane\cite{Muhlbauer-MnSiSKX2009}. Small angle neutron scattering (SANS) observes this triple-$\mathbf{Q}$ state as a six-fold magnetic Bragg pattern in reciprocal space (Fig.~\ref{Fig1}). The orientation of the SKX with respect to the crystal lattice is determined by the magneto-crystalline coupling, which has three contributions: (1) intrinsic spin-orbit coupling (SOC), (2) lattice anisotropy arising as a consequence of the discrete spin lattice, and (3) anisotropic demagnetization effect, which depends on the sample shape, and as we show below is negligible here. In the SKX, the plane containing all $\mathbf{Q}$s aligns perpendicular to $\mathbf{H}$. The lattice anisotropy further breaks the rotational symmetry of the triple $\mathbf{Q}$s within this plane, aligning the SKX to a preferred crystal axis, with one of its principal axes along either [100]- or [110]-axis\cite{Muhlbauer-MnSiSKX2009,Munzer-FeSi_CoSKX}.

For a detailed discussion, we define two angles that denote the orientation of $\mathbf{H}$ and the SKX with respect to the underlying crystalline lattice [see Fig.~\ref{Fig1}(b)]. $\theta$ is the angle between the [001] direction and $\mathbf{H}$, which rotates in the plane spanned by [010] and [001]. Moreover, $\psi$ is the angle between [100] and $\mathbf{Q}_1$, the closest of the three helical propagation vectors that reside in the plane defined by [100] and [0$q_y$$q_z$] perpendicular to $\mathbf{H}$ and that we refer to as the SKX coupling vector. $\psi$=0 indicates $\mathbf{Q}_1$$\parallel$[100], whereas $\psi$=90$\textordmasculine$ implies $\mathbf{Q}_1$$\parallel$[0$q_y$$q_z$].

For $\mathbf{H}$ along the highest-symmetry directions [001] ($\theta$=0), the plane selected for the SKX exhibits four-fold rotation symmetry and contains both axes [100] and [110], respectively. Here the magneto-crystalline coupling determines whether $\mathbf{Q}_1$ pins to [100] or [110]. However, in each case the four-fold symmetry allows for two degenerate SKX domains as shown in Fig.~\ref{Fig1}(c, d). This is observed in Fe$_{1-x}$Co$_x$Si, where two SKX domains coexist, with $\mathbf{Q}_1$ along [100] and [010], respectively [Fig.~\ref{Fig1}(c)]\cite{Munzer-FeSi_CoSKX}. For $\mathbf{H}$$\parallel$[011] ($\theta$=45$\textordmasculine$) the rotation symmetry is only two-fold and the SKX orientation is uniquely determined with $\mathbf{Q}_1$$\parallel$[01$\overline{1}$], as illustrated in Fig.~\ref{Fig1}(f) and observed in SANS experiments on MnSi\cite{Muhlbauer-MnSiSKX2009}. The orientation of the SKX with respect to the lattice for intermediate angle $0$$<$$\theta$$<$45$\textordmasculine$  remains an open question. We expect that $\mathbf{Q}_1$ remains pinned to [100] because the SKX plane only contains a single [100] easy axis [Fig.~\ref{Fig1}(e)]. However, as $\mathbf{H}$ is rotated, the magneto-crystalline coupling of the SKX varies, thus affecting SKX stability and coupling. This is partly supported by an AC susceptibility study which shows that the area of the SKX phase decreases for $\mathbf{H}$$\parallel$[110] compared to $\mathbf{H}$$\parallel$[100]\cite{Bauer-MnSiPhase}. We reveal the magnetic-field-angle dependence of the magneto-crystalline coupling via RUS, which measures sound waves scattering  by magnetic fluctuations.


High-quality single crystalline MnSi was grown by the Bridgman method. The sample was carefully polished into a parallelepiped along (001) axes with the dimensions 1.446$\times$0.485$\times$0.767 mm$^3$. AC susceptibility measurements reveal $T_c$=28.7 K at zero field. A schematic diagram of the RUS setup is shown in Fig.~\ref{Fig1}(b). More details about sample characterization, experimental and data analysis can be found in our earlier works\cite{LuoY-MnSiRUS1,Fobes-MnSiStrain}. The external magnetic field $\textbf{H}$ was rotated within the $\textbf{yz}$-plane from $\theta$=0 to 50$\textordmasculine$ in a step of 5$\textordmasculine$. All through the paper the results are presented as a function of applied magnetic field ($H$), and the internal field can be estimated by $B$$\approx$0.89 $H$ due to the demagnetization correction\cite{Aharoni-Demag}.


\begin{figure}[t]
\vspace*{-17pt}
\hspace*{-13pt}
\includegraphics[width=9.5cm]{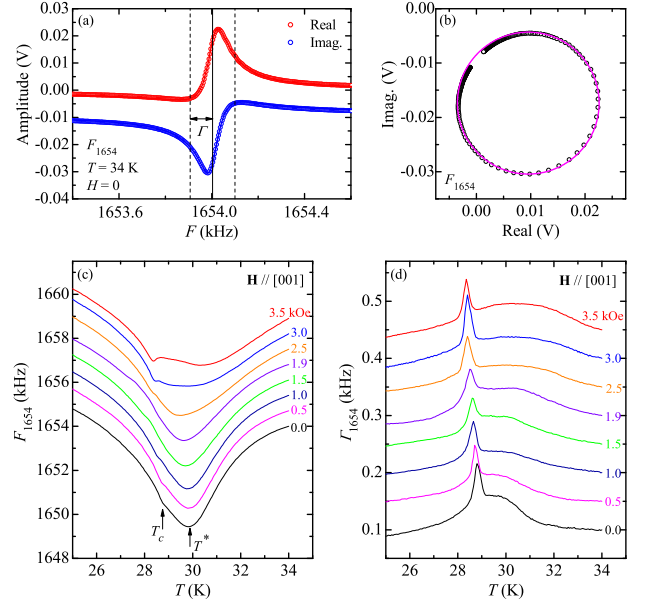}
\vspace*{-27pt}
\caption{\label{Fig2} (a) The resonant peak at $\sim$1654 kHz, denoted as $F_{1654}$. The solid lines represent Lorentzian-resonance fits\cite{Migliori-RUS}. ${\it\Gamma}$ is the characteristic width of the peak. (b) Imaginary vs. real parts of the resonance $F_{1654}$. (c) and (d) show the temperature dependence of $F_{1654}$ and ${\it\Gamma}_{1654}$, respectively, at various magnetic fields $\textbf{H}$$\parallel$[001]. Curves are vertically offset for clarity.}
\end{figure}

Figure \ref{Fig2}(a) shows the raw data of a representative resonant peak taken at 34 K in the absence of a magnetic field. The real part (in-phase) and imaginary part (out-of-phase) of the resonance track a circle nicely as shown in Fig.~\ref{Fig2}(b), and can be well fitted to a Lorentzian function\cite{Migliori-RUS}, allowing precise determination of the peak position $F$=1654.02 kHz and a characteristic peak width ${\it\Gamma}$=0.10 kHz. Hereafter we denote this resonance as $F_{1654}$. Because ${\it\Gamma}$ is proportional to the sound attenuation of the elastic mode, it provides a measure of spin fluctuations and their dissipation through the sound wave scattering\cite{Migliori-RUS}.

In Fig.~\ref{Fig2}(c), we present the temperature dependence of $F_{1654}$ for various applied fields $\mathbf{H}$$\parallel$[001]. Three prominent features can be identified for $H$=0. (\rmnum{1}) The profile of $F_{1654}(T)$ matches that of $C_{11}(T)$\cite{Petrova-PRB2015,LuoY-MnSiRUS1}. In a previous study, we showed that this resonance is dominated by the elastic modulus $C_{11}$\cite{LuoY-MnSiRUS1}. (\rmnum{2}) $F_{1654}$ decreases upon cooling and minimizes at around $T^*$=29.8 K, which reflects softening of elastic moduli in this temperature range. (\rmnum{3}) Below $T^*$, $F_{1654}$ turns up and exhibits an inflection point at $T_c$=28.7 K. The window between $T^*$ and $T_c$ corresponds to the fluctuation-disordered (FD) region [Fig.~\ref{Fig1}(a)], where the abundance of critical fluctuations of the helical order parameter suppresses both the correlation length and the mean-field helical phase transition and results in a Brazovskii-type fluctuation-induced first-order transition at $T_c$\cite{Brazovskii-1975, Janoschek-MnSi2013, Kindervater-MnSi2014}. Under an applied field, $T_c$ decreases monotonically. In contrast, the minimum position of $F_{1654}$ initially decreases but then increases for $H$$\geq$2.5 kOe due to the appearance of the PPM phase.

Figure \ref{Fig2}(d) plots ${\it\Gamma}_{1654}$ as a function of $T$ for $\mathbf{H}$$\parallel$[001]. At zero field, ${\it\Gamma}_{1654}(T)$ mimics the temperature dependent specific heat\cite{Bauer-MnSiC,Stishov-MnSiCT}, showing a broad shoulder around $T^*$ followed by a sharp peak at $T_c$. With increasing magnetic field the peak slowly moves to lower temperature and sharpens near 3 kOe, characteristic of approaching a tri-critical point\cite{Nii-MnSiElastic}. The shoulder above $T_c$ extends to higher temperature for $H$$\geq$2.5 kOe as the spins are polarized, consistent with $F_{1654}(T)$.

\begin{figure}[t]
\vspace*{-46 pt}
\hspace*{-20pt}
\includegraphics[width=11cm]{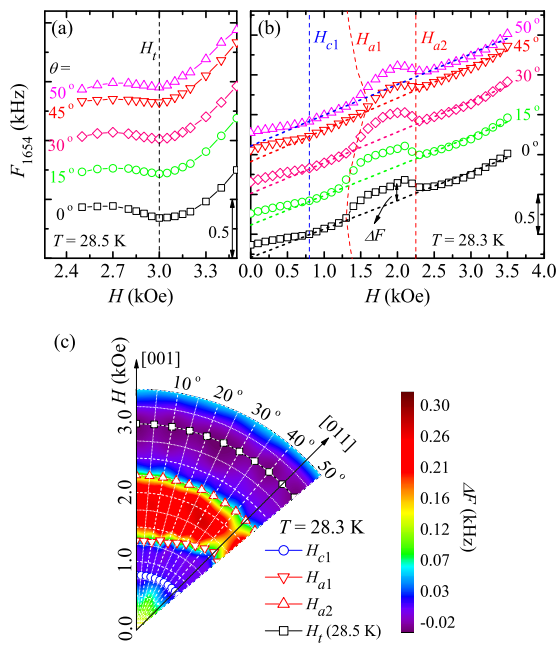}
\vspace*{-90pt}
\caption{\label{Fig3} Magnetic field dependence of $F_{1654}$ at $T$=28.5 K (a) and 28.3 K (b) for selected field orientations. (c) Polar contour plot of $\Delta F(H, \theta)$ at 28.3 K. $\Delta F$ is obtained by subtracting a linear background from $F_{1654}$, see panel (b).}
\end{figure}

Before we discuss the angular dependence of $F_{1654}$ and ${\it\Gamma}_{1654}$, we demonstrate that anisotropy due to demagnetization effect is negligible. At $T$=28.5 K [Fig.~\ref{Fig3}(a)] the valley position at $H_t$ (that is a signature of approaching the tri-critical region\cite{Nii-MnSiElastic}) remains essentially unchanged from $\theta$=0 to 50$\textordmasculine$. More evidence to exclude an anisotropic demagnetization effect comes from the nearly isotropic $H_{c1}$ (the transition field from HM to CO) and $H_{a2}$ (the upper boundary between SKX and CO) as shown in Fig.~\ref{Fig3}(b) measured at 28.3 K.

\begin{figure}[t]
\vspace*{-5pt}
\hspace*{-5pt}
\includegraphics[width=9.5cm]{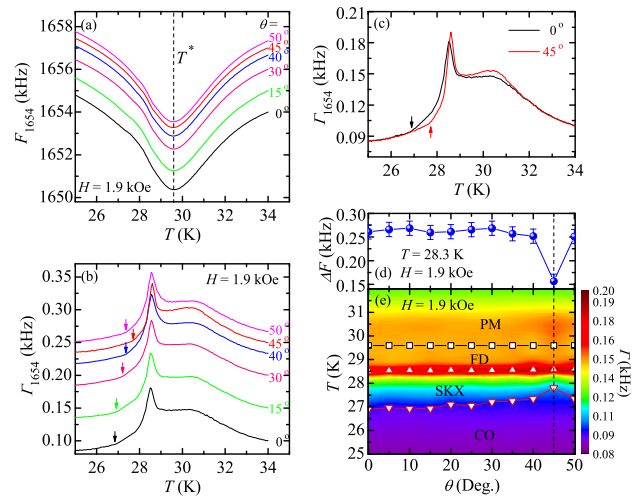}
\vspace*{-10pt}
\caption{\label{Fig4} Temperature dependence of $F_{1654}$ (a) and ${\it\Gamma}_{1654}$ (b) at various $\theta$ with fixed magnetic field strength $H$=1.9 kOe. The curves are vertically shifted for clarity. The dashed line in (a) signifies $T^*$, and the arrows in (b) denote the low-temperature boundary of the SKX phase. (c) Comparison of ${\it\Gamma}_{1654}$ for $\theta$=0 and 45$\textordmasculine$. (d) $\Delta F$ at 1.9 kOe as a function of $\theta$. (e) Angular dependent phase diagram at $H$=1.9 kOe constructed by a false-color plot of ${\it\Gamma}(\theta, T)$. }
\end{figure}

In order to investigate the angular changes of $F_{1654}(H)$, we subtract a linear background determined by values at $H$=1.0 and 2.5 kOe for each curve [see Fig.~\ref{Fig3}(b)]. The obtained $\Delta F(H, \theta)$ is displayed in a polar contour plot in Fig.~\ref{Fig3}(c). This demonstrates that the width of SKX phase gradually narrows as field rotates from [001] to [011] as is also evident from Fig.~\ref{Fig3}(b) where the critical field $H_{a1}$ (the lower boundary between CO and SKX) increases with $\theta$ up to $\theta$=45$\textordmasculine$. The SKX phase widens again when $\theta$$>$45$\textordmasculine$. In addition, the magnitude of $\Delta F$ in the SKX phase, a measure of magneto-crystalline coupling\cite{LuoY-MnSiRUS1}, decreases substantially near $\theta$=45$\textordmasculine$ [Fig.~\ref{Fig3}(c) and Fig.~\ref{Fig4}(d)]. Taken together these observations suggest that the magneto-crystalline coupling between the SKX and crystalline lattice weakens {\it abruptly} when magnetic field is parallel to [011].

The iso-field temperature dependence of $F_{1654}$ and ${\it\Gamma}_{1654}$ for different $\theta$ are shown in Fig.~\ref{Fig4}(a) and (b), respectively. $H$=1.9 kOe was selected because it allows access to the SKX phase for all $\theta$. The position of the minimum in $F_{1654}(T)$, denoted with $T^*$, is rather isotropic, but the behavior of ${\it\Gamma}_{1654}(T)$ is anisotropic: (\rmnum{1}) Below $T_c$, we observe an inflection in ${\it\Gamma}_{1654}(T)$ denoted by arrows in Fig.~\ref{Fig4}(b). This inflection point coincides with the lower edge of the SKX phase, cf. Fig.~\ref{Fig1}(a) for $\mathbf{H}$$\parallel$[001] and Ref.~\cite{Bauer-MnSiPhase} for $\mathbf{H}$$\parallel$[011]. The inflection occurs at 26.9 K for $\theta$=0, gradually moves to 27.34 K as $\theta$ increases to 40$\textordmasculine$, and abruptly jumps to 27.8 K for $\theta$=45$\textordmasculine$. (\rmnum{2}) The shoulder above $T_c$ in ${\it\Gamma}_{1654}(T)$ is a plateau for $\theta$$\leq$40$\textordmasculine$ but forms a broad peak for $\theta$=45$\textordmasculine$ [Fig.~\ref{Fig4}(c)], indicative of more fluctuations accumulating in the FD regime for this direction. We note that the behavior observed for $\theta$$<$40$\textordmasculine$ is recovered when $\mathbf{H}$ continues to rotate away from 45$\textordmasculine$. These results are summarized in a contour plot of ${\it\Gamma}_{1654}(\theta, T)$ in Fig.~\ref{Fig4}(e).

Our results suggest an abrupt change of the magneto-crystalline coupling when approaching $\theta$=45$\textordmasculine$. A possible explanation may be a spontaneous switch of the SKX orientation, as the depicted in smaller magneto-crystalline coupling and SKX extension [Fig.~\ref{Fig3}(c)] and changes in attenuation [Fig.~\ref{Fig4}(c)]. Once two SKX configurations pinned to distinct crystallographic axes are energetically competing, an increase of spin fluctuations is expected, similar to the increased attenuation we observe in the FD regime. This will affect the SKX stability as reflected in the substantial shrinkage of the SKX phase at the low-$T$ and $H$ boundaries as well as the reduced $\Delta F$ observed in our experiments.

\begin{figure}[t]
\vspace*{-15pt}
\hspace*{-13pt}
\includegraphics[width=9.5cm]{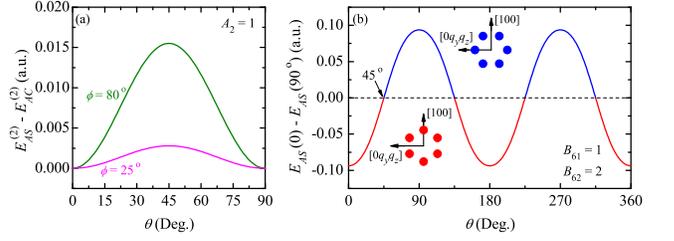}
\vspace*{-20pt}
\caption{\label{Fig5} (a) $\theta$ dependence of $E_{AS}^{(2)}$$-$$E_{AC}^{(2)}$ simulated with $\phi$=25$\textordmasculine$ and 80$\textordmasculine$. The curves are vertically shifted. (b) The difference of SKX coupling potential, $E_{AS}$($\psi$=0)$-$$E_{AS}$($\psi$=90$\textordmasculine$), as a function of $\theta$. The inset sketches denote SKX patterns pinned to $\psi$=0 (red) and 90$\textordmasculine$ (blue), respectively. }
\end{figure}

In the following we present the Ginzburg-Landau theory that supports this intuitive picture. For details, see \textbf{Supplemental Materials (SM)}\cite{SM}. We consider the phase boundary between the conical and SKX phases, and determine how the phase boundary changes when rotating the magnetic field. The leading term in the free energy determining the \emph{anisotropic} magneto-crystalline coupling arises from SOC and is given by
\begin{equation}
E_{A}^{(2)}(\theta)=\int\frac{A_2}{2}[(\partial_xS_x)^2+(\partial_yS_y)^2+(\partial_zS_z)^2]d^3\mathbf{r},
\label{Eq1}
\end{equation}
where $\mathbf{S}$=($S_x$,$S_y$,$S_z$) is the magnetization field. We take $\theta$=0 as a reference where the free energies of the SKX ($E_{AS}^{(2)}$) and CO ($E_{AC}^{(2)}$) phases are the same, and calculate their energy difference when $\theta$ changes
\begin{equation}
E_{AS}^{(2)}-E_{AC}^{(2)}=\frac{A_2 Q^2S^2}{128}\sin^2\phi[1-\cos4\theta],
\label{Eq2}
\end{equation}
where $\phi$$\sim$$\arccos(H/H_s)$ is the conical canting angle with $H_s$ being the saturation magnetic field at which all spins are polarized. A comparison of the numerical results for $\phi$=25$\textordmasculine$ and 80$\textordmasculine$ is shown in Fig.~\ref{Fig5}(a). Here $E_{AS}^{(2)}$$-$$E_{AC}^{(2)}$ is non-negative, indicating that the CO phase is energetically more favorable than the SKX phase as the field rotates. Moreover, $E_{AS}^{(2)}$$-$$E_{AC}^{(2)}$ changes faster for larger $\phi$ (lower phase boundary of SKX and CO phases). The angular phase diagram in Fig.~\ref{Fig3}(c) well reproduces these features. For the lower boundary, the SKX phase is invaded by the CO phase and substantially shrinks as $\theta$ increases. The higher boundary remains essentially unchanged, but a closer examination of Fig.~\ref{Fig3}(b-c) yields that the width of the transition broadens notably near $\theta$=45$\textordmasculine$, therefore, it is likely that the CO phase slightly wins at the higher boundary.

We also calculated the locking potential $E_{AS}(\psi)$ responsible for fixing the orientation of the SKX. The SKX has 6-fold rotation symmetry in the plane perpendicular to $\mathbf{H}$; therefore, the relevant terms in free energy functional has the order of $Q^6$ \cite{Muhlbauer-MnSiSKX2009,Munzer-FeSi_CoSKX},
\begin{equation}
\begin{aligned}
E_{AS}(\psi)&\simeq\displaystyle{\sum_{i=1}^{3}}\mathbf{S}(\mathbf{Q_i})\cdot\mathbf{S}(-\mathbf{Q_i})[B_{61}(Q_{ix}^6+Q_{iy}^6+Q_{iz}^6)\\
&~~+B_{62}(Q_{ix}^4Q_{iy}^2+Q_{iy}^4Q_{iz}^2+Q_{iz}^4Q_{ix}^2)].
\end{aligned}
\label{Eq3}
\end{equation}
Indeed, the minimum of $E_{AS}(\psi)$ occurs at either $\psi$=0 or 90$\textordmasculine$ \cite{SM}. We compare the energy for the two orientations of SKX with $\psi$=0 and 90$\textordmasculine$, and calculate their difference, $E_{AS}(0)$$-$$E_{AS}(90\textordmasculine)$, in the \textbf{SM}\cite{SM}. For any positive $B_{61}$ and $B_{62}$, $E_{AS}(0)$$-$$E_{AS}(90\textordmasculine)$ is negative at small $\theta$, but undergoes a sign change at a critical angle $\theta_c$, implying that the SKX switches from $\psi$=0 to 90$\textordmasculine$. In particular, when $B_{62}/B_{61}$=2, this sign change occurs exactly at $\theta_c$=45$\textordmasculine$ [Fig.~\ref{Fig5}(b)], in good agreement with our experimental results. It is worthwhile to mention that for $\mathbf{H}$$\parallel$[011], $\psi$=90$\textordmasculine$ overlaps with [01$\overline{1}$] [see Fig.~\ref{Fig1}(f)], consistent with previous SANS experiments\cite{Muhlbauer-MnSiSKX2009}. This calculation also predicts that for $\mathbf{H}$$\parallel$[001], the SKX prefers to pin at [100] rather than [110]. Our angular RUS measurements agree with this prediction because no anomaly is observed near $\theta$=0. Naturally, for this case degenerate multi-domains as shown in Fig.~\ref{Fig1}(c) may arise.

Summarizing, crucial aspects of our work are: (\rmnum{1}) We establish RUS as a tool to not only give information about elasticity but also about dissipation mechanisms of the atomic lattice with high precision, sensitive to detect changes in the magneto-crystalline coupling that determines the orientation of the SKX (Note \cite{note}). (\rmnum{2}) In analogy to superconducting vortices\cite{Das-CeCoIn5Vortex,Lin-VortexStrain}, the magneto-crystalline coupling in the SKX phase enables one to manipulate the skyrmions by external stress. (\rmnum{3}) Our work also reveals how the stability of the SKX phase changes as a function of its orientation with respect to the crystalline lattice. Our previous work also demonstrates in addition that the dynamics of skyrmions, such as the depinning transition in the presence of an electric current, can be inferred from the RUS measurements\cite{LuoY-MnSiRUS1}. Taken together this provides important information for future applications based on skyrmions.


In conclusion, we studied the field-rotation effect on the skyrmion lattice in MnSi using RUS. Abrupt changes are observed in elastic moduli and attenuation when the field is parallel with the [011] crystallographic direction, which suggests a reorientation of the SKX pattern, with a different signature in the FD and SKX regions. The phenomenon is associated with an anisotropic magneto-crystalline coupling potential between SKX and atomic lattices. Our work demonstrates that this coupling potential can be modeled in excellent agreement with experiment using a relatively simple phenomenological Ginzburg-Landau theory and, in turn, that the current understanding of SKX in cubic B20 materials is remarkably complete.

{\it Note added.} During the preparation of this manuscript, we noticed that similar phenomena were observed by SANS in MnSi\cite{Bannenberg-SKXOrien}.


We thank F. F. Balakirev for technical support, and F. Ronning and M. Garst for insightful conversations. Work at Los Alamos was funded by Laboratory Directed Research and Development (LDRD) program. ZL acknowledges a Director's Postdoctoral Fellowship supported through the LANL LDRD program. JB and AM acknowledge funding support from EFRC-Actinides (BES).


%

\newpage

\setcounter{table}{0}
\setcounter{figure}{0}
\setcounter{equation}{0}
\renewcommand{\thefigure}{S\arabic{figure}}
\renewcommand{\thetable}{S\arabic{table}}
\renewcommand{\theequation}{S\arabic{equation}}
\onecolumngrid

\newpage

\begin{center}
{\bf \large
{\it Supplemental Material:}\\
Anisotropic magnetocrystalline coupling of the skyrmion lattice in MnSi
}
\end{center}

\small
\begin{center}
Yongkang Luo$^{1*}$\email{mpzslyk@gmail.com}, Shi-Zeng Lin$^{1}$, D. M. Fobes$^{1}$, Zhiqi Liu$^{1}$, E. D. Bauer$^{1}$, J. B. Betts$^{1}$, A. Migliori$^{1}$, J. D. Thompson$^{1}$, M. Janoschek$^{1}$, and B. Maiorov$^{1\dag}$\email{maiorov@lanl.gov}\\
$^1${\it Los Alamos National Laboratory, Los Alamos, New Mexico 87545, USA;}\\

\date{\today}
\end{center}

\normalsize
\vspace*{15pt}
In this \textbf{Supplemental Material (SM)}, we provide details about the Ginzburg-Landau theory to understand the effect of magneto-crystalline coupling on the skyrmion crystal (SKX) phase and orientation of the SKX.

\section{\textbf{SM \Rmnum{1}: F\lowercase{ree energy density}}}

We expand the free energy in terms of magnetization field \textbf{S} and the ordering wavevector \textbf{Q}. We expand to the quartic order in \textbf{S} and 6th order in \textbf{Q}. The total free energy functional is
\begin{equation}
\mathcal{F}=\mathcal{F}_m+\mathcal{F}_{\mathrm{ANI}},
\label{EqS1}
\end{equation}
with
\begin{equation}
\mathcal{F}_m=\frac{1}{2}\alpha\textbf{S}^2+\frac{1}{4}\beta\textbf{S}^4+\frac{1}{2}J_2(\nabla \textbf{S})^2+D\textbf{S}\cdot(\nabla\times\textbf{S})+\frac{J_4}{2}(\nabla^2 \textbf{S})^2+\frac{J_6}{2}(\nabla^3 \textbf{S})^2-\textbf{S}\cdot\textbf{H},
\label{EqS2}
\end{equation}
\begin{equation}
\begin{aligned}
\mathcal{F}_{\mathrm{ANI}}=&\frac{C_4}{6}(S_x^4+S_y^4+S_z^4)+\frac{A_2}{2}[(\partial_xS_x)^2+(\partial_yS_y)^2+(\partial_zS_z)^2]+\frac{A_4}{2}[(\partial_x^2S_x)^2
+(\partial_y^2S_y)^2+(\partial_z^2S_z)^2]\\
&+\frac{A_6}{2}[(\partial_x^3S_x)^2+(\partial_y^3S_y)^2+(\partial_z^3S_z)^2]+\frac{B_4}{2}[(\partial_x^2\textbf{S})^2+(\partial_y^2\textbf{S})^2
+(\partial_z^2\textbf{S})^2]+\frac{B_{61}}{2}[(\partial_x^3\textbf{S})^2+(\partial_y^3\textbf{S})^2+(\partial_z^3\textbf{S})^2]\\
&+\frac{B_{62}}{2}[(\partial_x^2\partial_y\textbf{S})^2+(\partial_y^2\partial_z\textbf{S})^2+(\partial_z^2\partial_x\textbf{S})^2]\\
\end{aligned}
\label{EqS3}
\end{equation}
The skyrmion orientation is determined by $\mathcal{F}_{\mathrm{ANI}}$, which we will consider as a small perturbation. Formally the Ginzburg-Landau free energy functional is valid in the temperature region close to $T_c$, where the order parameter $\mathbf{S}(\mathbf{r})$ is small. Nevertheless, qualitative behavior of the system can also be inferred by extending the theory to the low temperature region.

\begin{figure*}[b]
\vspace*{-40pt}
\includegraphics[width=15cm]{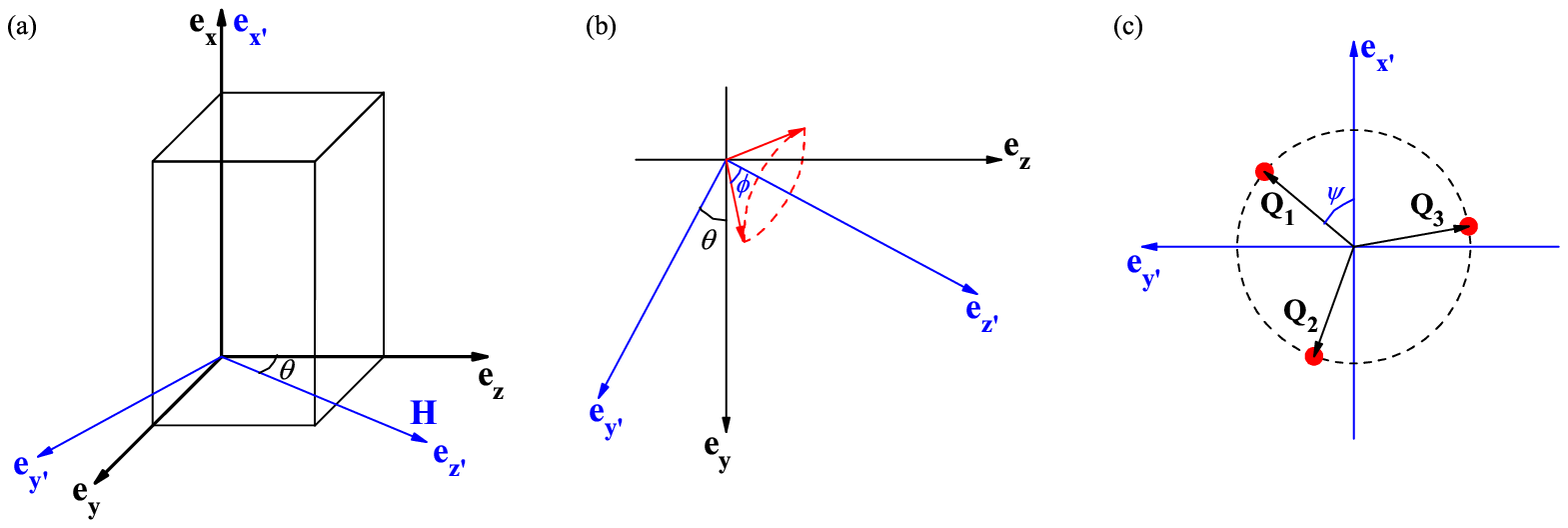}
\vspace*{-10pt}
\caption{\label{FigS1} (a) Schematic sketch of field rotation within the $\mathbf{yz}$-plane, $\mathbf{H}$$\parallel$$\mathbf{e_{z'}}$. (b) Conical phase where $\phi$ is the conical spanning angle. (c) Skrymion phase, with $\psi$ the angle between $\mathbf{Q_1}$ and $\mathbf{e_{x'}}$. }
\end{figure*}

We rotate the external magnetic field within \textbf{yz}-plane, with an angle $\theta$ from the \textbf{z}-axis, see Fig.~\ref{FigS1}(a). For convenience, we introduce a new rotated frame,
\begin{eqnarray}
\begin{aligned}
\mathbf{e_{x'}}&=~\mathbf{e_{x}},\\
\mathbf{e_{y'}}&=0\mathbf{e_{x}}+\cos\theta \mathbf{e_{y}}-\sin\theta \mathbf{e_{z}},\\
\mathbf{e_{z'}}&=0\mathbf{e_{x}}+\sin\theta \mathbf{e_{y}}+\cos\theta \mathbf{e_{z}}.
\end{aligned}
\label{EqS4}
\end{eqnarray}

In the conical phase, the wavevector $\mathbf{Q}$ is parallel to $\mathbf{H}$, viz.
\begin{equation}
\mathbf{Q}=0\mathbf{e_{x'}}+0\mathbf{e_{y'}}+Q\mathbf{e_{z'}}=Q(0, \sin\theta, \cos\theta).
\label{EqS5}
\end{equation}
The magnetization field $\mathbf{S}$ can be written as:
\begin{equation}
\begin{aligned}
\mathbf{S}&= S\sin\phi\cos(\mathbf{Q}\cdot\mathbf{r}) \mathbf{e_{x'}}+ S\sin\phi\sin(\mathbf{Q}\cdot\mathbf{r}) \mathbf{e_{y'}}+ S\cos\phi \mathbf{e_{z'}}\\
&=S\sin\phi\cos(\mathbf{Q}\cdot\mathbf{r})\mathbf{e_{x}}+S[\sin\phi\sin(\mathbf{Q}\cdot\mathbf{r})\cos\theta+\cos\phi\sin\theta]\mathbf{e_{y}}+
S[\cos\phi\cos\theta-\sin\phi\sin(\mathbf{Q}\cdot\mathbf{r})\sin\theta]\mathbf{e_{z}},
\end{aligned}
\label{EqS6}
\end{equation}
where $\phi$ is the conical canting angle between $\mathbf{S}$ and $\mathbf{H}$ [cf. Fig.~\ref{FigS1}(b)].

The SKX can be considered as a superposition of three-$\mathbf{Q}$ helices, where all the three $\mathbf{Q}$s are perpendicular to magnetic field $\mathbf{H}$, as depicted in Fig.~\ref{FigS1}(c). First, we need to write down the three wavevectors:
\begin{eqnarray}
\begin{aligned}
\mathbf{Q_1}&=Q\cos\psi\mathbf{e_{x'}}+Q\sin\psi\mathbf{e_{y'}}=Q[\cos\psi, \sin\psi\cos\theta, -\sin\psi\sin\theta],\\
\mathbf{Q_2}&=Q\cos(\psi+2\pi/3)\mathbf{e_{x'}}+Q\sin(\psi+2\pi/3)\mathbf{e_{y'}}=Q[\cos(\psi+2\pi/3), \sin(\psi+2\pi/3)\cos\theta, -\sin(\psi+2\pi/3)\sin\theta],\\
\mathbf{Q_3}&=Q\cos(\psi-2\pi/3)\mathbf{e_{x'}}+Q\sin(\psi-2\pi/3)\mathbf{e_{y'}}=Q[\cos(\psi-2\pi/3), \sin(\psi-2\pi/3)\cos\theta, -\sin(\psi-2\pi/3)\sin\theta],~~~~~~~~
\end{aligned}
\label{EqS7}
\end{eqnarray}
where $\psi$ is defined as the angle between $\mathbf{Q_1}$ and $\mathbf{e_x}$. The corresponding helical magnetization fields are:
\begin{equation}
\begin{aligned}
\mathbf{S_1}=&-S_{\perp}\sin\psi\sin(\mathbf{Q_1}\cdot\mathbf{r})\mathbf{e_{x}}+S_{\perp}[\cos\psi\sin(\mathbf{Q_1}\cdot\mathbf{r})\cos\theta+S_{\perp}\cos(\mathbf{Q_1}\cdot\mathbf{r})\sin\theta+S_0\sin\theta]\mathbf{e_y}\\
&+[S_{\perp}\cos(\mathbf{Q_1}\cdot\mathbf{r})\cos\theta+S_0\cos\theta-S_{\perp}\cos\psi\sin(\mathbf{Q_1}\cdot\mathbf{r})\sin\theta]\mathbf{e_{z}},\\
\mathbf{S_2}=&-S_{\perp}\sin(\psi+2\pi/3)\sin(\mathbf{Q_2}\cdot\mathbf{r})\mathbf{e_{x}}+S_{\perp}[\cos(\psi+2\pi/3)\sin(\mathbf{Q_2}\cdot\mathbf{r})\cos\theta+S_{\perp}\cos(\mathbf{Q_2}\cdot\mathbf{r})\sin\theta+S_0\sin\theta]\mathbf{e_y}\\
&+[S_{\perp}\cos(\mathbf{Q_2}\cdot\mathbf{r})\cos\theta+S_0\cos\theta-S_{\perp}\cos(\psi+2\pi/3)\sin(\mathbf{Q_2}\cdot\mathbf{r})\sin\theta]\mathbf{e_{z}},\\
\mathbf{S_3}=&S_{\perp}\sin(\psi-2\pi/3)\sin(\mathbf{Q_3}\cdot\mathbf{r})\mathbf{e_{x}}-S_{\perp}[\cos(\psi-2\pi/3)\sin(\mathbf{Q_3}\cdot\mathbf{r})\cos\theta+S_{\perp}\cos(\mathbf{Q_3}\cdot\mathbf{r})\sin\theta+S_0\sin\theta]\mathbf{e_y}\\
&-[S_{\perp}\cos(\mathbf{Q_3}\cdot\mathbf{r})\cos\theta+S_0\cos\theta-S_{\perp}\cos(\psi-2\pi/3)\sin(\mathbf{Q_3}\cdot\mathbf{r})\sin\theta]\mathbf{e_{z}},
\label{EqS8}
\end{aligned}
\end{equation}
where $S_{\perp}$=$S\sin\phi$, and $S_0$=$S\cos\phi$.  The total magnetization field for the SKX is:
\begin{equation}
\begin{aligned}
\mathbf{S}=&\mathbf{S_1}+\mathbf{S_2}+\mathbf{S_3}\\
=&[-S_{\perp}\sin\psi\sin(\mathbf{Q_1}\cdot\mathbf{r})-S_{\perp}\sin(\psi+2\pi/3)\sin(\mathbf{Q_2}\cdot\mathbf{r})+S_{\perp}\sin(\psi-2\pi/3)\sin(\mathbf{Q_3}\cdot\mathbf{r})]\mathbf{e_x}\\
&+[S_{\perp}\cos\psi\cos\theta\sin(\mathbf{Q_1}\cdot\mathbf{r})+S_{\perp}\sin\theta\cos(\mathbf{Q_1}\cdot\mathbf{r})+S_{\perp}\cos(\psi+2\pi/3)\cos\theta\sin(\mathbf{Q_2}\cdot\mathbf{r})+S_{\perp}\sin\theta\cos(\mathbf{Q_2}\cdot\mathbf{r})\\
&~~~~-S_{\perp}\cos(\psi-2\pi/3)\cos\theta\sin(\mathbf{Q_3}\cdot\mathbf{r})-S_{\perp}\sin\theta\cos(\mathbf{Q_3}\cdot\mathbf{r})+S_0\sin\theta]\mathbf{e_y}\\
&+[S_{\perp}\cos\theta\cos(\mathbf{Q_1}\cdot\mathbf{r})-S_{\perp}\cos\psi\sin\theta\sin(\mathbf{Q_1}\cdot\mathbf{r})+S_{\perp}\cos\theta\cos(\mathbf{Q_2}\cdot\mathbf{r})-S_{\perp}\cos(\psi+2\pi/3)\sin\theta\sin(\mathbf{Q_2}\cdot\mathbf{r})\\
&~~~~-S_{\perp}\cos\theta\cos(\mathbf{Q_3}\cdot\mathbf{r})+S_{\perp}\cos(\psi-2\pi/3)\sin\theta\sin(\mathbf{Q_3}\cdot\mathbf{r})+S_0\cos\theta]\mathbf{e_z}.~~~~
\label{EqS9}
\end{aligned}
\end{equation}

\section{\textbf{SM \Rmnum{2}: P\lowercase{hase boundaries of conical phase} \lowercase{and} SKX}}
We compare the energy of the conical phase and SKX phase when one rotates the magnetic field. At the phase boundary for $\theta$=0 or $\mathbf{H}$$\parallel$$\mathbf{e}_z$, these two phases have the same energy. As as field is rotated, the energy change is a consequence of the  contribution in $\mathcal{F}_{\mathrm{ANI}}$. Close to the transition temperature $T_c$, $S$ is small and in the long wavelength limit, the dominant contribution is given by the $A_2$ term in Eq. \eqref{EqS3}.

The energy for the conical phase is
\begin{equation}
E_{AC}^{(2)}=\frac{A_2Q^2S^2}{16}\sin^2\phi(1-\cos4\theta),
\label{EqS10}
\end{equation}
and the energy for the SKX phase is
\begin{equation}
E_{AS}^{(2)}=\frac{A_2Q^2S^2}{128}\sin^2\phi(33-9\cos4\theta),
\label{EqS11}
\end{equation}
At $\theta$=$0$, the conical phase and SKX have the same energy, meaning there is an energy difference in $\mathcal{F}_m$ between these two phases. Taking the energy difference in $\mathcal{F}_m$ into account, the energy difference between the SKX and conical phase is
\begin{equation}
\Delta E_A(\theta)=E_{AS}^{(2)}-E_{AC}^{(2)}=\frac{A_2Q^2S^2}{128}\sin^2\phi(1-\cos4\theta).
\label{EqS12}
\end{equation}
The canting angle is $\cos\phi$$\sim$$H/H_s$, with $H_s$ being the saturation field at which all moments are polarized. In Fig.~\ref{Fig5}(a), we plot the $\theta$ dependence of $\Delta E_A$ for $\phi$=25$\textordmasculine$ and 80$\textordmasculine$. For both $\phi$=25$\textordmasculine$ and 80$\textordmasculine$, $\Delta E_A$ becomes positive when $\theta$ increases, implying that the CO phase is energetically more favorable. Another observation is that $\Delta E_A(\theta)$ increases much faster when $\phi$=80$\textordmasculine$. Our angular phase diagram in Fig.~\ref{Fig3}(c) well reproduces these features. For the lower boundary, the SKX phase is invaded by the CO phase and substantially narrows as $\theta$ increases. The higher boundary remains essentially unchanged, but by a closer examination to Fig.~\ref{Fig3}(b-c), one finds that the width of the transition broadens especially near $\theta$=45$\textordmasculine$. It, therefore, is likely that the CO phase slightly wins at the higher boundary but better experimental resolution is needed to clarify it.

\section{\textbf{SM \Rmnum{3}: $\theta$ \lowercase{dependent} SKX \lowercase{pattern}}}

We turn to the $\psi$ dependence of the coupling potential of SKX lattice to atomic lattice. Previous works by S. M\"{u}hlbauer {\it et al.} have revealed that the easy axis of the SKX intensity pattern is determined by the terms of the order of $Q^6$\cite{Muhlbauer-MnSiSKX2009,Munzer-FeSi_CoSKX}, which are given by
\begin{equation}
\begin{aligned}
E_{AS}^{(61)}&\equiv B_{61}\displaystyle{\sum_{\mathbf{Q_i}}}[(Q_{ix}^6+Q_{iy}^6+Q_{iz}^6)\mathbf{S}(\mathbf{Q_i})\cdot\mathbf{S}(-\mathbf{Q_i})]\\
&=\frac{B_{61}Q^6S_{\perp}^2}{2}[(\cos^6\psi+\sin^6\psi\cos^6\theta+\sin^6\psi\sin^6\theta)+(\psi\rightarrow\psi+2\pi/3)+(\psi\rightarrow\psi-2\pi/3)].
\label{EqS13}
\end{aligned}
\end{equation}

\begin{equation}
\begin{aligned}
E_{AS}^{(62)}&\equiv B_{62}\displaystyle{\sum_{\mathbf{Q_i}}}[(Q_{ix}^4Q_{iy}^2+Q_{iy}^4Q_{iz}^2+Q_{iz}^4Q_{ix}^2)\mathbf{S}(\mathbf{Q_i})\cdot\mathbf{S}(-\mathbf{Q_i})]\\
&=\frac{B_{62}Q^6S_{\perp}^2}{2}[(\cos^4\psi\sin^2\psi\cos^2\theta+\sin^6\psi\cos^4\theta\sin^2\theta+\sin^4\psi\cos^2\psi\sin^4\theta)+~~~~~~~~~~~~~~~~\\
&~~~~~~~~~~~~~~~~~~(\psi\rightarrow\psi+2\pi/3)+(\psi\rightarrow\psi-2\pi/3)].
\label{EqS14}
\end{aligned}
\end{equation}
We define $E_{AS}(\psi)$$\simeq$$E_{AS}^{(61)}(\psi)$$+$$E_{AS}^{(62)}(\psi)$.

\begin{figure*}[ht]
\vspace*{-10pt}
\hspace*{-15pt}
\includegraphics[width=19cm]{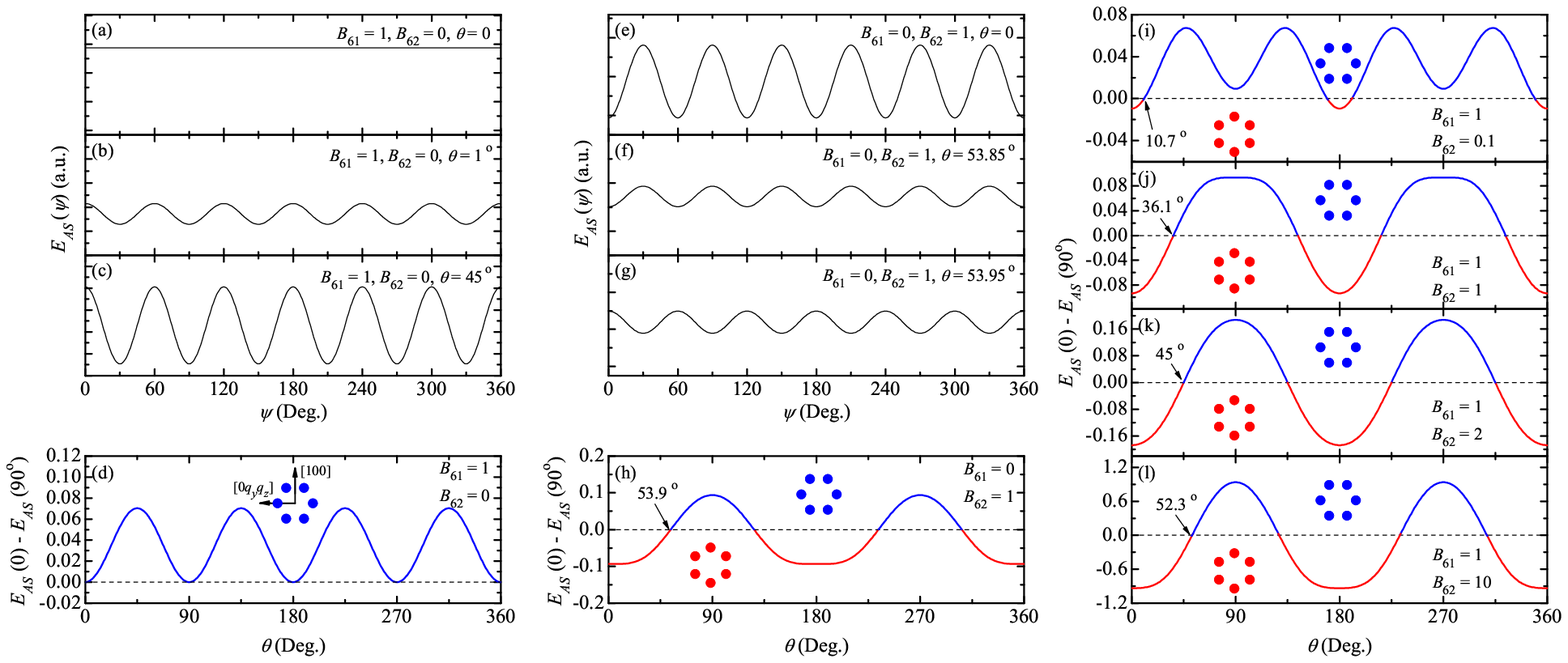}
\vspace*{-30pt}
\caption{\label{FigS2} SKX coupling. The first column (a-d) are calculated with $B_{61}$=1, and $B_{62}$=0; the second column (e-h) are calculated with $B_{61}$=0, and $B_{62}$=1; the third column (i-l) are calculated for different $B_{62}$ with fixed $B_{61}$=1. (a-c) and (e-g) show the $\psi$ dependence of $E_{AS}$ at various fixed field angle $\theta$. The energy difference between $\psi$=0 and $\psi$=90$\textordmasculine$, noted $E_{AS}(0)$-$E_{AS}(90$\textordmasculine$)$, is plotted in panels (d) and (h), respectively. (i-l) $E_{AS}(0)$$-$$E_{AS}(90$\textordmasculine$)$ for different ratios of $B_{62}$/$B_{61}$, showing that a sign change occurs at different values of $\theta_c$.}
\end{figure*}

We start from setting $B_{62}$=0, so $E_{AS}(\psi)$ is purely $E_{AS}^{(61)}(\psi)$. The calculations shown in Fig.~\ref{FigS2}(a-c) manifest that $E_{AS}(0)$ is always larger than $E_{AS}(90\textordmasculine)$, except for $\theta$=0 (or 90$\textordmasculine$) in which situation $E_{AS}$ is $\psi$-independent. This means that the $B_{61}$ term will always force the SKX lattice to pin at $\psi$=90 $\textordmasculine$ only except when magnetic field is parallel to [001] or [010], see Fig.~\ref{FigS2}(d).

We now turn to another extreme where $B_{61}$=0 and $B_{62}$=1. Interestingly, in this case, $E_{AS}(0)$$-$$E_{AS}(90~\textordmasculine)$ is negative for small field angle $\theta$, but undergoes a sign change at $\theta_c$=53.9$\textordmasculine$ [Fig.~\ref{FigS2}(e-g)], suggesting an abrupt switch from $\psi$=0 SKX pattern to $\psi$=90$\textordmasculine$ pattern, see Fig.~\ref{FigS2}(h).

The combination of $B_{61}$ and $B_{62}$ terms renders tunability of critical angle $\theta_c$. Fig.~\ref{FigS2}(i-l) display $E_{AS}(0)$$-$$E_{AS}(90\textordmasculine)$ as a function of $\theta$ calculated with various ratios of $B_{62}/B_{61}$. As expected, the larger $B_{62}/B_{61}$ yields the larger $\theta_c$ with the latter saturating at 53.9$\textordmasculine$ in the limit of $B_{62}/B_{61}$$\rightarrow$$+$$\infty$. In particular, when $B_{62}/B_{61}$=2, the SKX pattern switch occurs exactly at 45$\textordmasculine$ [cf. Fig.~\ref{FigS2}(k)], which well explains our experimental results. We should emphasize that it is impossible for SKX pattern to switch at 45$\textordmasculine$ if $B_{61}$ and $B_{62}$ are of opposite signs.

\begin{figure*}[ht]
\vspace*{-10pt}
\hspace*{-0pt}
\includegraphics[width=14cm]{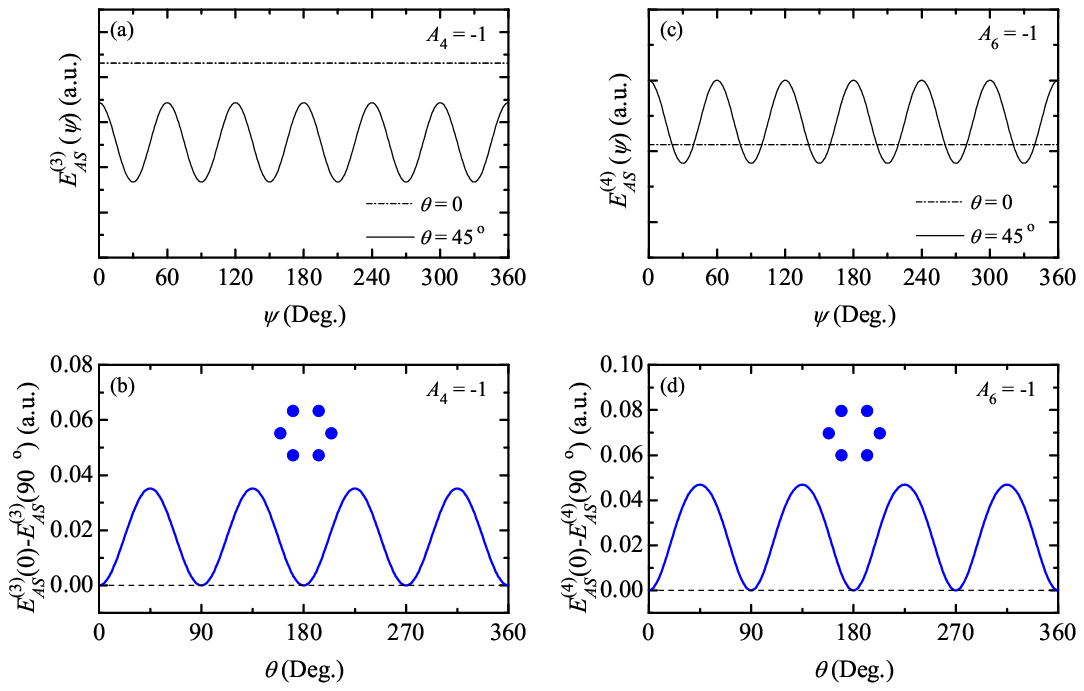}
\vspace*{-15pt}
\caption{\label{FigS3} $\psi$ dependencies of $E_{AS}^{(3)}$ (a) and $E_{AS}^{(4)}$ (c), calculated with $A_{4}$=$-$1 and $A_{6}$=$-$1, respectively. Panels (b) and (d) display $E_{AS}^{(3)}(0)-E_{AS}^{(3)}(90\textordmasculine)$ and $E_{AS}^{(4)}(0)-E_{AS}^{(4)}(90\textordmasculine)$. In both cases, SKX are preferably pinned to $\psi$=90$\textordmasculine$.}
\end{figure*}

It should be pointed out that the $B_{61}$ and $B_{62}$ terms are not the only terms in Eq.~(\ref{EqS3}) that show $\psi$-dependence. The following contributions
\begin{equation}
\begin{aligned}
E_{AS}^{(3)}&\equiv A_4\displaystyle{\sum_{\mathbf{Q_i}}}[Q_{ix}^4S_x(\mathbf{Q_i})S_x(-\mathbf{Q_i})+Q_{iy}^4S_y(\mathbf{Q_i})S_y(-\mathbf{Q_i})+Q_{iz}^4S_z(\mathbf{Q_i})S_z(-\mathbf{Q_i})]\\
&=\frac{A_4Q^4S_{\perp}^2}{4}\{[\sin^2\psi\cos^2\psi(\cos^2\psi+\sin^2\psi\sin^6\theta+\sin^2\psi\cos^6\theta)+\sin^4\psi\sin^2\theta\cos^2\theta]+~~~~~~~~~~~~~~~~~~~\\
&~~~~~~~~~~~~~~~~~[\psi\rightarrow\psi+2\pi/3]+[\psi\rightarrow\psi-2\pi/3] \},
\label{EqS15}
\end{aligned}
\end{equation}

\begin{equation}
\begin{aligned}
E_{AS}^{(4)}&\equiv A_6\displaystyle{\sum_{\mathbf{Q_i}}}[Q_{ix}^6S_x(\mathbf{Q_i})S_x(-\mathbf{Q_i})+Q_{iy}^6S_y(\mathbf{Q_i})S_y(-\mathbf{Q_i})+Q_{iz}^6S_z(\mathbf{Q_i})S_z(-\mathbf{Q_i})]\\
&=\frac{A_6Q^6S_{\perp}^2}{4}\{[\sin^2\psi\cos^2\psi(\cos^4\psi+\sin^4\psi\sin^8\theta+\sin^4\psi\cos^8\theta)+\sin^6\psi(\sin^6\theta\cos^2\theta+\sin^2\theta\cos^6\theta)]+\\
&~~~~~~~~~~~~~~~~~[\psi\rightarrow\psi+2\pi/3]+[\psi\rightarrow\psi-2\pi/3]\},
\label{EqS16}
\end{aligned}
\end{equation}
also exhibit $\psi$ dependence. However, they behave similarly as $E_{AS}^{(61)}(\psi)$ when both $A_4$ and $A_6$ are negative [Fig.~\ref{FigS3}], and thus can be absorbed by the $B_{61}$ term. Therefore, the discussions above remain valid, but the condition that leads to SKX pattern switches at 45$\textordmasculine$ now becomes $6B_{61}Q^2$$-$$3A_4$$-$$4A_6Q^2$=$3B_{62}Q^2$.

\end{document}